# Nonlinear Depolarization of Light in Optical Communication Fiber

Lothar Moeller

*Abstract*— We report the experimental observation of a novel transmission phenomenon in optical long-haul communication systems. Un-polarized ASE depolarizes via nonlinear fiber interactions a cw laser light during their co-propagation which leads to small but measurable ultra-fast polarization state fluctuations at the fiber output. We provide a phenomenological approach and a theory that qualitatively corroborates our experimental results. One of our major findings suggests that the applicability of the often used Manakov equation needs to be scrutinized for highly accurate studies of nonlinear polarization state evolutions in noisy environments. The described phenomenon leads to a qualitatively different microscopic understanding of nonlinear light propagation in fiber and can contribute towards an explanation for today's commonly perceived gap between simulated and experimentally obtained system performance in optical data transmission.

## I. INTRODUCTION

MODERN optical fiber communications, the basis of all backbone networks, enable global long-reach and high-capacity data exchange like the WWW. State-of-the-art systems provide 10's Tb/s capacity per fiber over transpacific distances without electrical signal regeneration[1]. This success has been facilitated by powerful simulation-supported system development that carefully balances linear and nonlinear (NL) signal impairments in fibers. Nevertheless, the undersea communication industry continues to strive for improved accuracy in system design, as any performance gain corresponds to longer unrepeated system spans and higher transport capacities, hence more cost-effective solutions.

However, a noticeable discrepancy in channel capacity between measurable performance and predictions from the most advanced simulation tools remains, which attracts significant research in propagation modelling mostly focusing on pure optical effects[2] but lately also considering weak acousto-optical interactions[3,4]. Here we discuss a recently observed transmission phenomenon that we refer as NL depolarization of light (NLDP) and which can narrow down this discrepancy. Unpolarized optical noise rapidly changes via the fiber's Kerr nonlinearity the state of polarization (SOP) from a fully polarized cw light by inducing antisymmetric phase noise in both of its orthogonal polarization states. These fluctuations become resolvable with a new generation of high-speed polarimeters and do not average out over wide noise bandwidths, but grow with propagation distance.

Various aspects of NL polarization evolution in fiber have been studied extensively[5,6,7,8] often by applying a set of differential equations in Stokes space[9], derived from the Manakov equation[10], or by directly using NL differential equations[11] under the assumption of deterministic[12] or at least quasi-deterministic[13] SOPs. In contrast, we utilize the coupled NL Schrodinger equations for birefringent media and describe NLDP as interference between a signal with deterministic SOP and noise with truly random polarization, ideally possessing a zero-degree of polarization (DOP). This ansatz leads to qualitatively diverging results relative to those predicted by the Manakov equation and also is experimentally verified.

We start with a phenomenological approach that provides a first qualitative understanding of NLDP based on experimental data taken from a commercial cable system. Although somewhat limited by restrictions seen in field work, this experiment is sufficiently meaningful to prove the existence of NLDP in commercial networks. Thereafter described fiber test beds for laboratory usage advantageously allow parametric studies of NLDP and emulate nearly ideal system conditions conducive to mathematical analysis. We then introduce a NLDP theory derived by finding analytic solutions for the coupled Schrodinger equations in the weak nonlinear region. Subsequent remarks give an outlook of the NLDP impact on the development of system simulation tools and established theoretical limits for the channel capacity of fiber links.

## II. OBSERVATION OF NONLINEAR DE-POLARIZATION IN A DEPLOYED CABLE SYSTEM – THE MEASUREMENT PRINCIPLE

A regular undersea communication network, connects two landing stations at North and South America. Prior to its commissioning for commercial operation in August 2017, the cable was available for the following experimental work.

Undersea communication links usually bear customized designs with individual optical characteristics. Here we provide a generic system view, detailed to a degree which a qualitative understanding of our measurements requires. The link supplies an end-to-end throughput capacity of 10.7 Tb/s per direction delivered over 4.3 THz system bandwidth and consists of 6 about 10.556 km long fiber pairs, resembling a quasi-periodic concatenation of 165 spans, each of them formed by an optical repeater and standard single mode fiber (SSMF) of ~64 km length (Fig.1a). The repeaters comprise mainly an industry-typical two-stage Erbium Doped Fiber Amplifier (EDFA) design followed by passive equalization filters and insert ~18 dBm power into the transmission fiber. At the North American station, the output of one fiber is amplified and looped back to South America by launching it into the input of a return link (optical loopback).

The key idea behind our experiment is to launch completely

Lothar Moeller is with SubCom, Eatontown, NJ 07724, USA (lmoeller@subcom.com).

unpolarized ASE (here after referred to as loading, DOP<1%) from South America together with a weak, fully polarized and spectrally separated cw light (probe) to visualize NL crosstalk in the latter's receive SOP based on a comparative test.

At launch, the loading overlaps the repeater bandwidth except for a narrow gap of ~100 GHz, centered at 193.4 THz (Fig.1a) where the probe, generated by an external cavity laser (ECL), resides. During transmission the repeaters spectrally confine the loading which carries by far more power than the probe (~ -7.2 dBm at repeater output) to avoid Brillouin scattering or modulation instability. A simple passive optical filter network combines probe and loading on the transmit side; whereas on the receive side, a filter (FWHM ~27 GHz) selects the probe from the pre-amplified fiber output. Gain and filter width are chosen such that the polarimeter detects the probe at high optical signal-to-noise-ratio (OSNR) and with sufficient optical power (~ -4 dBm). A detailed analysis yields signal-ASE beat noise as the dominant distortion in the recording of Stokes parameters (ASE-ASE beat noise and thermal noise can be neglected). Our commercially available polarimeter samples and digitalizes at 14 bit the outputs of its four photodetectors every 10 ns. Signal post-processing computes the differential of two consecutive normalized Stokes vectors and represents its magnitude as SOP speed histogram with 1024 bins, all 48.82 krad/s wide.

Any noise in the detection process (ASE, A/D quantization, thermal noise) cause lateral motions of the Stokes vector on the Poincare sphere and artificial SOP speed. Even a hypothetical constant SOP of the probe on the receive side will appear in a histogram with non-zero width at least due to the omnipresent signal-ASE beat noise. We refer to this noise-induced SOP speed (NISS) as an artifact since improving the OSNR at the polarimeter input would reduce the width of its histogram. However, probe power constraints, inevitable added repeater noise during transmission, and practical limitations on tighter filtering yield OSNR levels which result in artificially broadened SOP speed histograms that partially obscure NLDP. Instead we demonstrate fast SOP changes due to NLDP using a comparison technique. We contrast the receive SOP speed distributions from the transmitted probe with one from a reference signal that possesses equal power, equal ASE, and equal OSNR but bypasses the cable (back-to-back signaling, btb). This reference is obtained by superimposing the transmitter signal with the noise output of the transmission link and launched via a short fiber jumper directly into the receiver (dashed path, Fig.1a). A switch toggles between both paths during a pairing of the features from probe and reference whose matching within ~0.1 dB near the laser line (Fig.1b) is verified by an optical spectrum analyzer (OSA). The SOP speed histograms (Fig.1c), recorded at identical polarimeter settings and constituted by about 10 GSamples, visualize the NLDP impact on the probe by its significantly expanded distribution. To ensure its diverse width is not caused by adjustment errors, we increase the ASE level by ~5 dB (noise boost) through adding additional noise to the reference on the receive side. Indeed, the corresponding SOP speed histogram broadens compared to the btb curve, but stays narrower than the distribution obtained for the probe. These results indicate faster SOP changes in the transmitted signal as can be explained by the magnitude of any artifacts. In our understanding, during transmission the unpolarized loading induces rapid polarization-dependent phase modulation onto the probe leading to SOP fluctuations. But as later theoretically discussed, this phase noise (~10 MHz spectral width) is below the OSA's optical resolution (~4 GHz).

The spectral features of our ECL, a linewidth < 50 kHz and negligible wavelength drift and RIN, allow to abstract away the analysis from the two linear depolarization processes polarization mode dispersion (PMD) and polarization dependent loss (PDL) as intrinsic cause for NLDP. NL polarization rotation[9,11,12] based effects, governed by the Manakov equation, can be ruled out as well. A 24 h SOP speed recording (at strongly reduced polarimeter bandwidth) quantifies environmental impact on the system such as temperature drift, stress, twist, or thunderstorms; but mainly cable motion on the ocean floor to < 20 rad/s which is irrelevant for NLDP.

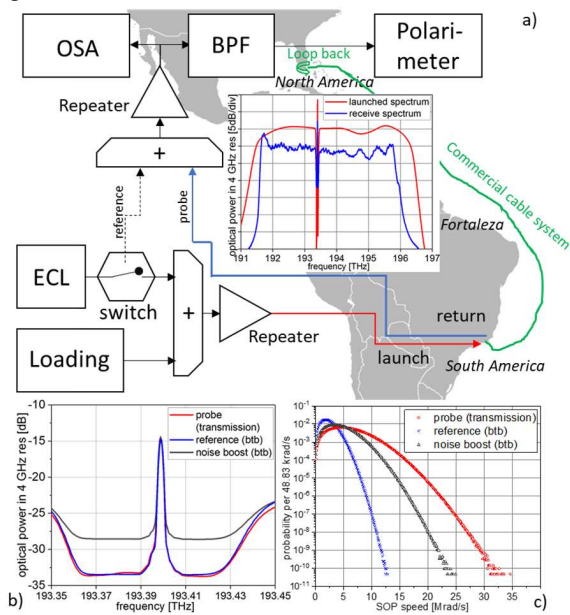

**Fig.1:** a) Transmitter and receiver designs for comparative polarimetry on a commercial cable system. A cw tone propagates together with unpolarized ASE from South America to North America and back. Its SOP speed is measured at the South American station and compared with a local reference (not transmitted through the cable, dashed path). b) matched optical spectra of the probe and reference, taken with an OSA, look similar but the corresponding SOP speed histograms (c) differ significantly and prove the existence of NLDP (Noise boost explained in text).

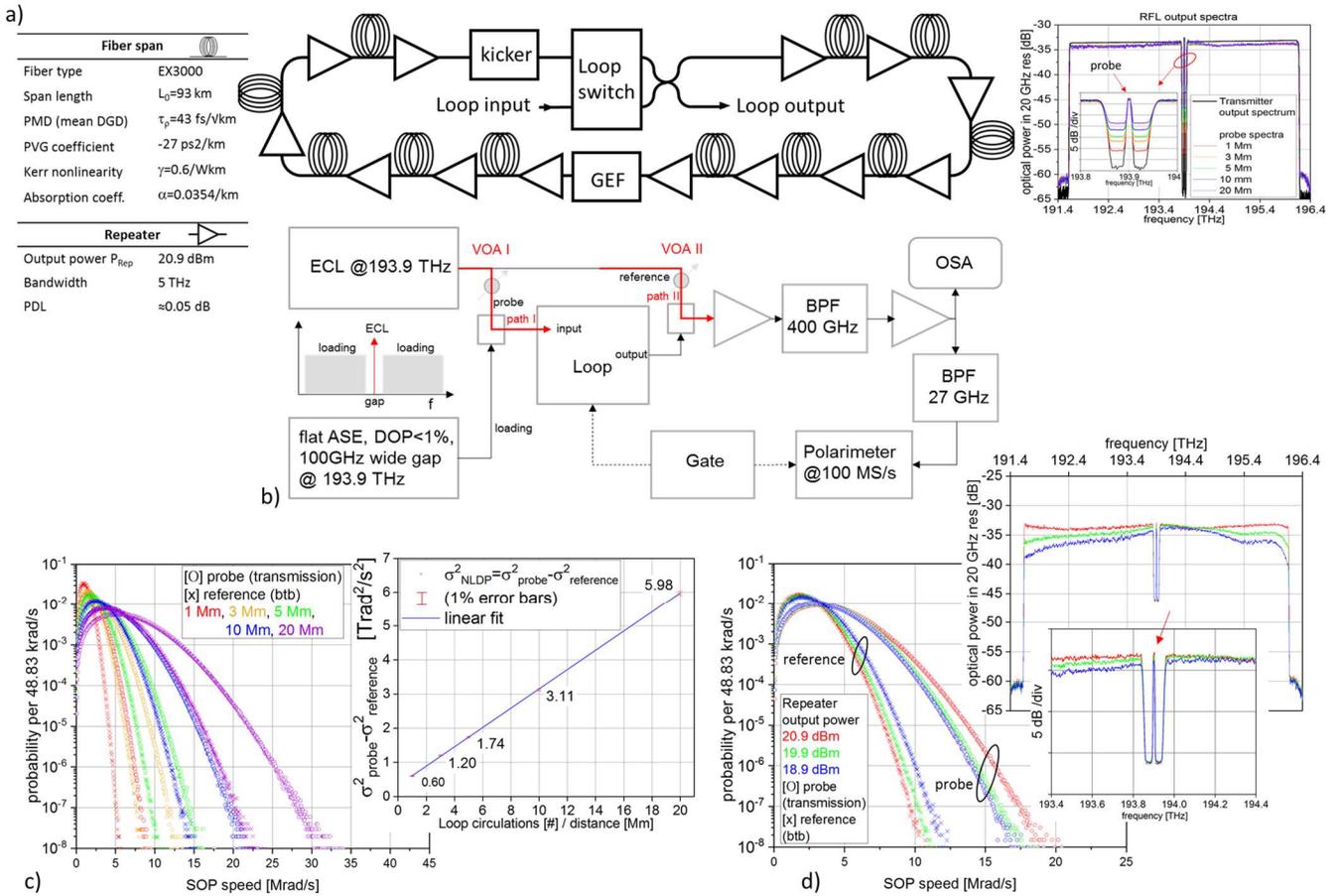

**Fig.2: Characterizing NLDP in a lab setup using a recirculating fiber loop.** a) RFL design with output spectra for different transmission distances and its fiber specifics. The spectrum of the loading stays relatively flat after transmission which eases a modelling of NLDP. b) experimental setup for comparative polarimetry using a RFL. The RFL allows emulating differently long transmission paths which impacts the magnitude of NLDP. The transmission length is determined by a central clock that synchronizes the gates of the RFL, the polarimeter, and the OSA. The SOP speeds of the probe after transmission and of the reference are analyzed using a high-speed polarimeter. c) visualization of the NLDP dependence on transmission distances; the histogram widths for probe and reference increase with transmission distance. The difference $\sigma^2_{NLDP}$ between the SOP speed variances of the probe and reference shows the NLDP magnitude and grows linearly with transmission distance. d) repeater power dependence of NLDP at 10 Mm transmission. The Kerr effect which mediates the cross-phase modulation between loading and probe is power dependent. Lowering the repeater output power reduces the width of the SOP speed histogram from the probe and deforms the spectral shape of the loading.

In all our experiments a polarization controller (not shown in Fig.1a) slowly scrambles the ECL's output SOP quasi-homogeneously across the Poincare sphere. This mitigates linear polarization effects along the cable with negligible bias on the histograms.

### III. QUANTIFYING NLDP IN A RECIRCULATING FIBER LOOP

#### A. Nonlinear De-Polarization vs. Transmission distance

It is reasonable to assume a scaling of the NLDP magnitude with the probe's transmission length and the loading power, as both determine the Kerr nonlinearity strength. However, installed cable systems are designed for fixed repeater output power operation. In contrast, parametric NLDP-studies become feasible with a lab test bed based on a recirculating fiber loop (RFL)[14]. Methodically identical to the field experiment, we polarimetrically compare a transmitted probe and a reference with equal OSNR and power.

The RFL comprises of 11 transmission spans each with an output power-adjustable repeater followed by SSMF (specs see Fig.2a). A loop switch passes the signal through the RFL and a gain equalizing filter (GEF) mitigates wavelength-dependent attenuation in the transmission path. A triggered polarization controller[15] (kicker) randomizes PMD and PDL inside the loop. After the signal has passed the kicker and before its return, the kicker is set to a different mode within a few ns to randomize the output SOP.

On the transmitter side (Fig.2b), the probe passes a variable optical attenuator (VOA) and enters via path I the RFL (VOA_I open, VOA_II blocked). An amplified filter cascade selects from the RFL output the probe prior to its detection by the polarimeter. To acquire the reference for specifying NISS (btb measurement), VOA_I blocks the access to path I while the reference reaches the receiver via path II. In both scenarios, loading enters the RFL via path I. We tune VOA_II to match the spectra from probe and reference within 0.1 dB across a bandwidth of approx. 50 GHz (OSA, Fig.2a). A central clock synchronizes the gates form loop switch, kicker, OSA, and polarimeter. It periodically activates the latter's data acquisition at 100 MS/s typically for a ~3 ms long interval per transmission. A programmable ASE source delivers unpolarized loading (DOP<1%) with flat spectrum (constant power density) across the repeater gain bandwidth of about 5 THz except for a gap (100 GHz wide, ~50 dB ASE suppression) centered at 193.9 THz to contain the probe. Here we chose a flat launch spectrum and a flat transfer function for the RFL, adjusted via its GEF (Fig.2a), to simplify the below given mathematical treatment of NLDP.

By varying the loop's timing gate, SOP speed histograms are recorded (Fig.2c) for distances of approx. 1023 km, 3069 km, 5115 km, 10,230 km, and 20,460 km further referred to as 1 Mm, 3 Mm, …, 20 Mm transmissions and correspond to 1, 3, 5, 10, and 20 loop circulations, respectively. The relating btb measurements show wider NISS histograms for longer transmissions as their receive OSNRs decay. SOP speed histograms after transmission broaden with propagation length due to NISS and NLDP. We hypothesize both as statistically independent processes, and visualize NLDP by subtracting from the probe's SOP speed variance the NISS from the reference (Fig.2c, inset). This quantity monotonically increases, indicating a NLDP growth with transmission distance. A linear fit reasonably resembles the measured NLDP variances as a function of the transmission length. Each recording consists of about 100 MSamples and was repeated multiple times to verify stable measurement readings. We found experimentally for the short-term reproducibility of all measured SOP speed variances a relative error smaller than 1%. Altering the nominal probe power of ~-5.2dBm at repeater output by ±3 dB has shown an insignificant dependence of $\sigma^2_{NLDP}$ on self-phase modulation based effects at 10 Mm transmission distance. Thus, modulation instability or Brillouin scattering, the known to be weakest NL fiber processes, can be ruled out as origin for NLDP.

### B. Nonlinear Depolarization vs Repeater Output Power

The repeater output power of our RFL can be controlled to some degree without strongly tilting their transfer functions. At maximum repeater output power (20.9 dBm into transmission fiber) and at 1 and 2 dB reduced strengths, we record SOP speed histograms over 10 Mm transmission distance following the same methodology as above. Increasing widths of the reference's NISS histograms with decreasing repeater power (Fig.2d) can be explained by receive OSNR depletion. To a good approximation, the ASE a repeater adds to an amplified signal is proportional to its gain. Since our repeaters operate in constant gain mode, altering the output power does not affect their ASE contributions. Consequently, lower repeater output power means lower probe power and OSNR after transmission (at constant ECL power) which boosts NISS. Remarkably, after transmission, and despite improved receive OSNR, the corresponding SOP speed histograms trend oppositely and widen with enhanced repeater power. This broadening stems from NLDP which even exceeds a theoretically expected narrowing of the histograms when only the enhanced receive OSNRs are considered as in case of the reference.

Other than in the NLPD vs transmission distance study, the received spectra deform at different test conditions (Fig.2d, inset). While around 194.4 THz the spectral density remains almost constant, it drops disproportionally at the spectral edges with reduced repeater power making a quantitative analysis more challenging. But the noise floor surrounding the probe stays relatively constant which supports previous OSNR argument.

A certain kind of GAWBS[3] can cause SOP fluctuations but these motions are power-independent. Thus, the observed repeater power dependence of the SOP speed allows us to rule out GAWBS as origin for NLDP.

### IV. MATHEMATICAL APPROACH TO NLDP

Here we develop simplified analytic calculus techniques to qualitatively describe NLDP. Modern telecom fibers are often modeled as concatenation of birefringent waveplates typically ranging in length between 10 m and 100 m and possessing randomly orientated principal axes relative to each other. Published models differ regarding assumed distributions for the plate birefringence and length. These features have only secondary meaning for our analysis as long as they reside within the parameter range of typical SSMF and are here not detailed further. The coupled NL Schrodinger (CNLS) equations[16,17] describe well the slowly varying envelope of an overall field $A^\Sigma_{x(y)}$ written as superposition of the weak probe $a_{x(y)}$ and the much stronger loading $A_{x(y)}$ within a single waveplate with principal axes aligned along the x-y coordinates of the lab frame[5]

$$\frac{\partial A^\Sigma_{x(y)}}{\partial z} + \beta_{1x}\frac{\partial A^\Sigma_{x(y)}}{\partial t} + \frac{j\beta_{2x(y)}}{2}\frac{\partial^2 A^\Sigma_{x(y)}}{\partial t^2} + \frac{\alpha}{2}A^\Sigma_{x(y)} =$$
$$j\gamma\frac{5}{6}\left(|A^\Sigma_{x(y)}|^2 + |A^\Sigma_{y(y)}|^2\right)A^\Sigma_{x(y)} + j\gamma\frac{1}{6}\left(|A^\Sigma_{x(y)}|^2 - |A^\Sigma_{y(x)}|^2\right)A^\Sigma_{x(y)}$$
$$+ j\gamma\frac{1}{3}A^{\Sigma*}_{x(y)}A^{\Sigma 2}_{y(y)}e^{-(+)j2\Delta\beta z} \quad (1)$$

with
$$A^\Sigma_{x(y)} = A_{x(y)} + a_{x(y)}; \quad |a_{x(y)}| \ll |A_{x(y)}| \quad (2)$$

where the wavelength-independent $\alpha$, $\beta_{1x(y)}$, $\beta_{2x(y)}$, and $\gamma$ determine the attenuation, the polarization-dependent group velocities and their dispersion coefficients, and the Kerr nonlinearity of the fiber, respectively. The modal birefringence $\Delta\beta$, induces fast oscillating but ineffective NL interference, known from discussions[18] about the Manakov-PMD equation, however ignorable in our analysis. The remaining part of the right side from Eq. (1) is composed of a symmetric and an anti-symmetric term of the loading's field describing independent weak NL interactions, treatable in a first order perturbation calculus. The weak probe power allows rearranging the perturbation into a symmetric subset governing the probe field motions

$$\frac{\partial a_{x(y)}}{\partial z} + \beta_1 \frac{\partial a_{x(y)}}{\partial t} + \frac{j\beta_2}{2}\frac{\partial^2 a_{x(y)}}{\partial t^2} + \frac{\alpha}{2}a_{x(y)} = j\gamma\frac{4}{3}\left(|A_{x(y)}|^2 + |A_{y(x)}|^2\right)a_{x(y)}, \quad (3)$$

where other terms (in $a^2_{x(y)}$, $a_{x(y)}{}^*$, $A_{x(y)} A_{y(x)}{}^*$, etc.) are left out as they insufficiently contribute, and where due to the limited perturbation bandwidth (see below) a polarization-independent group velocity $\beta_1$ replaces $\beta_{1x(y)}$. We represent the right sides of Eq. (3) by means of the undistorted loading field (0$^{\text{th}}$ o. approximation) whose components $A_{x(y)}^n$ form a frequency comb on an evenly spaced grid with an infinitesimal small angular frequency pitch $\omega$ (Fig.3a). Their amplitudes are scaled such that $|A_x^n|^2+|A_y^n|^2$ matches the loading's spectral power density around the frequency $\omega/2\pi n$ at launch (z = 0). Probe and loading are assumed to start at z=0 and their series' in 0$^{\text{th}}$ o. approximation read

$$A_{0x(y)} = \sum_{n=-N}^{N} A_{x(y)}^n e^{j(k_n z-n\omega t)} e^{-\frac{\alpha}{2}z}\sigma(z); \ a_{0x(y)} = a_{x(y)}^0 e^{-\frac{\alpha}{2}z}\sigma(z) \quad (4)$$
$$\text{with } A_{x(y)}^n = 0 \text{ for } n\in\{-N_u+1, N_u-1\}\ V\ n > N$$

where $k_n = \beta_1 n\omega + \frac{\beta_2}{2}(n\omega)^2$, t, z, and $\sigma$ stand for the propagation constant of a component at '$n\omega$', the time, the propagation distance, and the Heaviside function, respectively. The probe's asymptotic propagation behavior, expanded as superposition of its undistorted state and 1$^{\text{st}}$ o. perturbation

$$a_{x(y)}(z,t) \approx a_{0x(y)}(z,t) + a_{1x(y)}(z,t) = a_{0x(y)}(z,t) + \sum_{l,m} a_{1x(y)}^{l,m}(z,t), \quad (5)$$

follows by deriving from (3) the distortions at '$l\omega$'

$$ja_{1x(y)}^{l,m}(z\to\infty,t) \approx \frac{4}{3}\gamma e^{-\frac{\alpha}{2}z}e^{j(k_l z-l\omega t)}\mathcal{A}_{m,l}(j\beta_2 ml\omega^2-\alpha)^{-1}a_{0x(y)} \quad (6)$$
$$\text{with } \mathcal{A}_{m,l} = A_x^{m+\frac{l}{2}}A_x^{m-\frac{l}{2}*} + A_y^{m+\frac{l}{2}}A_y^{m-\frac{l}{2}*}$$

and summing them over '$l$' and '$m$'. Typically, $m \gg l$ holds and justifies the denominator's simplification Eq. (6). Without loss of generality but for the sake of an easy readable symmetric notation, we assume '$l$' as even integer. Exchanging '$l$' by '$-l$' results into an exact conjugate complex right side of Eq. (6) except for the quadratic term of the probe's propagation constant which stays small even after ~10 Mm transmission. Hence, the overall distortion appears as quasi pure phase noise.

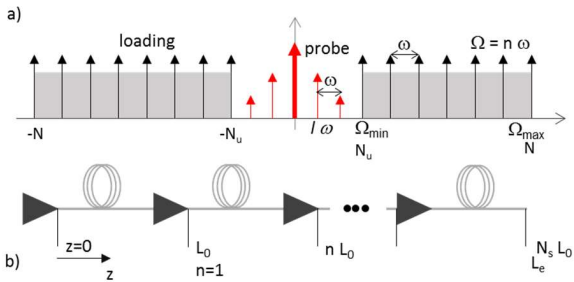

**Fig.3: | Key parameters of modelled signals and fiber link.**
a) Frequency combs defining fields from probe and loading; b) chain of $N_s$ identical transmission spans.

We extend the algorithm to a link consisting of $N_s$ identical spans of length $L_0$, by coherently adding at the link end $L_e$ all perturbations $a_{1x(y)}^{l,m}|_n$ generated in each span '$n$' using Eq. (6) under consideration of phase shifts chromatic dispersion induces to the phasors $A_{x(y)}^{m+\frac{l}{2}}, A_{x(y)}^{m-\frac{l}{2}*}$ (Fig.3b). We assume the same probe launch power at every span input, a linear propagation of all perturbations through their remaining link portion once they have been generated, and neglect repeater ASE.

$$a_{1x(y)}^{l,m}(L_e,t) = \sum_{n=1}^{N_s} a_{1x(y)}^{l,m}\big|_n (L_e,t)$$
$$\approx -j\frac{4}{3}\gamma e^{-\frac{\alpha}{2}L_0-jl\omega t}e^{j\beta_1 l\omega L_e}\mathcal{A}_{m,l}(j\beta_2 ml\omega^2-\alpha)^{-1}a_{0x(y)}\sum_{n=1}^{N_s} e^{j\beta_2 lm\omega^2 L_0 n} \quad (7)$$

The phasors obey independent stochastic processes with zero mean as $a_{1x(y)}^{l,m}(L_e,t)$ does. Its variance reads

$$\sigma_{l,m,x(y)}^2 = \langle a_{1x(y)}^{l,m}(L_e) a_{1x(y)}^{l,m}(L_e)^*\rangle \approx$$
$$\left(\frac{4\gamma}{3\alpha}\right)^2 N_s^2 e^{-\alpha L_0} \langle \mathcal{A}_{m,l}\mathcal{A}_{m,l}^*\rangle \frac{\left(1+\left(\frac{\beta_2 ml\omega^2}{\alpha}\right)^2\right)^{-1}}{1+\left(\frac{\beta_2 L_e ml\omega^2}{2}\right)^2}|a_{0x(y)}|^2 \quad (8)$$
$$\text{with } |l| \geq 1,$$

where $\langle\cdot\rangle$ denotes the phasors' ensemble average. In Eq. (8) a Lorentz curve with nearly the same area approaches the sum in Eq. (7). Summation over '$m$' yields the spectral properties of the perturbation that roll off slowly but are limited to about 10 MHz at a transmission distance of 10 Mm and typical system parameters. The total phase fluctuations caused by phasors '$m$' follows as

$$\sigma_{mx(y)}^2 \approx \frac{32}{9}\pi\left(\frac{\gamma}{\alpha}\right)^2 e^{-\alpha L_0} \frac{\langle |A_x^m|^2 |A_x^{m''}|^2 + |A_y^m|^2 |A_y^{m''}|^2 \rangle}{\beta_2 m\omega^2 L_0} N_s |a_{0x(y)}|^2, \quad (9)$$

where double primes indicate statistically independent phasors. For flat spectra as have been adjusted in Section III A., and a repeater power $P_{Rep}$ the spectral power density of the loading and the perturbation variance read

$$\langle |A_{x(y)}^m|^2\rangle = \langle |A_{x(y)}^{m''}|^2\rangle = \frac{1}{4}\frac{P_{Rep}}{(\Omega_{max}-\Omega_{min})}\omega \quad (10)$$

$$\sigma_{x(y)}^2 \approx \frac{8}{9}\pi\left(\frac{\gamma}{\alpha}\right)^2 e^{-\alpha L_0}\frac{1}{\beta_2 L_0}N_s\left(\frac{P_{Rep}}{(\Omega_{max}-\Omega_{min})}\right)^2 \ln\left(\frac{\Omega_{max}}{\Omega_{min}}\right)|a_{0x(y)}|^2. \quad (11)$$

The evenly strong and correlated phase noises in both polarizations do not explicitly vary the probe's SOP but slightly widen its spectrum and lead (in combination with fiber PMD) to extremely small 2$^{\text{nd}}$ order SOP fluctuations. The measurable SOP fluctuations result from anti-symmetric phase noise as will be discussed next. It may be helpful to envision the propagating beat signal of two phasors $A_{x(y)}^{m+\frac{l}{2}}, A_{x(y)}^{m-\frac{l}{2}*}$ as a wave, hereafter referred to as $G_{m\omega}$-wave, generating crosstalk.

The polarization-dependent sign of the perturbation terms on the right side of Eq. (12) leads to different phase noise in both principal axes, which manifests experimentally as NLDP

$$\frac{\partial a_{x(y)}}{\partial z} + \beta_{1x(y)}\frac{\partial a_{x(y)}}{\partial t} + \frac{j\beta_2}{2}\frac{\partial^2 a_{x(y)}}{\partial t^2} + \frac{\alpha}{2}a_{x(y)} = j\gamma\frac{2}{3}\left(|A_{x(y)}|^2 - |A_{y(x)}|^2\right)a_{x(y)}. \quad (12)$$

Solutions for Eq. (12) first model a single span as concatenation of many non-birefringent waveplates with individual lengths '$dz_i$' centered at position $z_i$ and incorporate in a second step their birefringent character ($\beta_{1x} \neq \beta_{1y}$). Each waveplate generates a perturbation whose asymptotic behavior is found for a single span as

$$\begin{pmatrix}a_{1x}^{l,m}\\a_{1y}^{l,m}\end{pmatrix}_{\substack{z\to\infty,t\\z_i,dz_i}} = \frac{-j2\gamma\alpha}{3}\frac{\mathcal{A}_{m,l}}{j\beta_2 ml\,{}^2-\alpha}e^{-\frac{\alpha}{2}z-\alpha z_i}e^{jk_l z-jl\omega t}\begin{bmatrix}1 & 0\\0 & -1\end{bmatrix}\begin{pmatrix}a_{0x}\\a_{0y}\end{pmatrix}_{\{z=0\}}dz_i \quad (13)$$

In our propagation model we incorporate the fiber birefringence, originating from $\beta_{1x} \neq \beta_{1y}$ by means of a Jones matrix that rotates the SOPs from probe and G-wave when traversing a waveplate. A Jones matrix of a consecutive waveplate $R_i$ shall be given by an angle $\xi_i$ defining its axes relative to those of $R_{i-1}$ and an accumulated phase difference $2\zeta_i$ in its principal states due to birefringence

$$\overline{\overline{R_1}} = \begin{bmatrix} e^{j\zeta_i}\cos\xi_i & e^{j\zeta_i}\sin\xi_i \\ -e^{-j\zeta_i}\sin\xi_i & e^{-j\zeta_i}\cos\xi_i \end{bmatrix}. \quad (14)$$

Next we prove that NL perturbations stemming from two different waveplates correlate to some degree which results into their partial coherent superposition at the fiber output and an unneglectable polarization-dependent phase noise essential for NLDP. The similarity between distortions from two waveplates '$i$' and '$i+q$' ($q \geq 1$) partially depends on the correlation of the NL operator $\mathcal{A}_{m,l}$ rated at both places

$$\left\langle \begin{pmatrix} a_{1x}^{l,m} \\ a_{1y}^{l,m} \end{pmatrix}_{\substack{z \to \infty, t \\ z_{i+q}, dz_{i+q}}} \middle| \begin{pmatrix} a_{1x}^{l,m} \\ a_{1y}^{l,m} \end{pmatrix}_{\substack{z \to \infty, t \\ z_i, dz_i}} \right\rangle \propto (\cos^2\xi - \sin^2\xi)\langle \mathcal{A}_{m,l}\mathcal{A}_{m,l}^*\rangle, \quad (15)$$

where $\xi$ stands for an effective angle of rotation caused by the interjacent Jones matrix product $\overline{\overline{R_\Pi}} = \prod_{i=n}^{i+q}\overline{\overline{R_n}}$, valued at the G-wave frequency ($= m\omega/2\pi$), and $\{|\}$ denotes the dot product. The perturbation vector Eq. (13) autocorrelates without the NL operator as

$$\left\{ \begin{pmatrix} a_{1x}^{l,m} \\ a_{1y}^{l,m} \end{pmatrix}_{\substack{z \to \infty, t \\ z_{i+q}, dz_{i+q}}} \middle| \begin{pmatrix} a_{1x}^{l,m} \\ a_{1y}^{l,m} \end{pmatrix}_{\substack{z \to \infty, t \\ z_i, dz_i}} \right\} \propto \left\{ \begin{bmatrix} 1 & 0 \\ 0 & -1 \end{bmatrix} \prod_{n=i}^{i+q} \overline{\overline{R_n}} \begin{pmatrix} a_{0x} \\ a_{0y} \end{pmatrix} \middle| \prod_{n=i}^{i+q} \overline{\overline{R_n}} \begin{bmatrix} 1 & 0 \\ 0 & -1 \end{bmatrix} \begin{pmatrix} a_{0x} \\ a_{0y} \end{pmatrix} \right\}$$
$$\propto (\cos^2\xi' - \sin^2\xi'), \quad (16)$$

where $\xi'$ denotes an effective angle of rotation caused by $\overline{\overline{R_\Pi}}$ at the frequency of the probe and $\binom{a_{0x}}{a_{0y}}$ is the probe's Jones vector at the input of $\overline{\overline{R_1}}$. Terms including $a_{0x}a_{0y}^*$ can be ignored. The theory of PMD statistics specifies the de-correlation from two SOPs[19] and provides a correlation for $\xi$ and $\xi'$

$$\langle(\cos^2\xi - \sin^2\xi)(\cos^2\xi' - \sin^2\xi')\rangle = \frac{1}{2}e^{-\frac{1}{2}\Delta\Omega^2\tau_\rho^2\Delta L}, \quad (17)$$

where $\Delta\Omega$, $\tau_\rho$, and $\Delta L$ are the angular frequency spacing between the two states, the mean fiber DGD per √length, and the fiber length, respectively. PMD effects within the bandwidth (~$l\omega$) of probe and G-wave are negligibly small.

Modern SSMF with low PMD allows assuming for the sake of simplicity a frequency-independent Jones matrix for the NL interaction length of a signal (~10 km). However, we consider the PMD-induced SOP misalignments from probe and G-waves after they have traversed at least one span. This yields the autocorrelation in time domain of the perturbation vector at '$l\omega$' caused by a $G_{m\omega}$-wave

$$\sigma_{NLDP}^2(L_e,\tau) = \left\langle \begin{pmatrix} a_{1x}^{l,m} \\ a_{1y}^{l,m} \end{pmatrix}_{\substack{L_e \\ t+\tau}} \middle| \begin{pmatrix} a_{1x}^{l,m} \\ a_{1y}^{l,m} \end{pmatrix}_{\substack{L_e \\ t}} \right\rangle$$

$$\approx \mathcal{H} \sum_{n,p=1}^{N_s,N_s} e^{j\beta_2 lm\omega^2 L_0(n-p)} \frac{1}{2} e^{-\frac{1}{2}(m\omega)^2\tau_\rho^2 L_0|n-p|} \quad (18)$$

$$= \mathcal{H} \sum_{n=-N_s}^{N_s} N_s \Lambda\left(\frac{n}{N_s}\right) e^{j\beta_2 lm\ ^2 L_0 n} \frac{1}{2} e^{-\frac{1}{2}(m\omega)^2\tau_\rho^2 L_0|n|}$$

$$\approx \mathcal{H} \left( \left( \sum_{n=-\infty}^{\infty} \frac{N_s}{2} e^{-\left(\frac{1}{2}(m\omega)^2\tau_\rho^2 L_0 + \frac{2}{N_s}\right)|n|} e^{j\beta_2 lm\ ^2 L_0 n} \right) \right) (19)$$

$$\approx \mathcal{H} N_s \frac{1}{\frac{1}{2}(m\omega)^2\tau_\rho^2 L_0 + \frac{2}{N_s}} \frac{1}{1 + \left(\frac{\beta_2 lm\ ^2 L_0}{\frac{1}{2}(m\omega)^2\tau_\rho^2 L_0 + \frac{2}{N_s}}\right)^2} \quad (20)$$

with $\mathcal{H} = e^{jl\omega\tau} \frac{4\gamma^2 e^{-\alpha L_0}}{9\alpha^2} \frac{\langle \mathcal{A}_{m,l}\mathcal{A}_{m,l}^*\rangle}{1+\left(\frac{\beta_2 ml}{\alpha}\ ^2\right)^2}\left(|a_{0x}|^2 + |a_{0y}|^2\right)$

Substituting the triangular function $\Lambda(\cdot)$ in Eq. (19) with an exponential term of equal integration area simplifies the summation as a Fourier integral Eq. (20). Considering electronic bandwidth limitations of the polarimeter (1$^{st}$ order low pass filter, 3-dB cut-off $\frac{\Omega_e}{2\pi} \approx 30\ MHz$) yields the autocorrelation of a perturbation caused by a $G_{m\omega}$-wave in the electrical domain

$$\sigma_{NLDP\atop m}^2(L_e,\tau)$$
$$\approx \left(\frac{2\gamma}{3\alpha}\right)^2 \frac{e^{-\alpha L_0 N_s}}{\frac{1}{2}(m\omega)^2\tau_\rho^2 L_0 + \frac{2}{N_s}} \sum_{l=-\infty}^{+\infty} e^{jl\omega\tau} \frac{\left(1+\left(\frac{\beta_2 ml\omega^2}{\alpha}\right)^2\right)^{-1}}{1+\left(\frac{\beta_2 lm\omega^2 L_0}{\frac{1}{2}(m\omega)^2\tau_\rho^2 L_0 + \frac{2}{N_s}}\right)^2} \frac{1}{1+\left(\frac{l\omega}{\Omega_e}\right)^2} \Xi_{m,l} \quad (21)$$

with $\Xi_{m,l} = \langle \mathcal{A}_{m,l}\mathcal{A}_{m,l}^*\rangle \left(|a_{0x}|^2 + |a_{0y}|^2\right)$.

For small $l\omega\tau$ and an $l$-independent $\Xi$ we develop $e^{jl\omega\tau} \cong 1 + jl\omega\tau - \frac{1}{2}(l\omega\tau)^2$ and obtain for the second moment in $\tau^2$

$$\sigma_{NLDP\atop m}^2(L_e,0) - \sigma_{NLDP\atop m}^2(L_e,\tau) =$$
$$\frac{2\gamma^2 e^{-\alpha L_0}\Omega_e^2}{9\beta_2 m\omega\alpha\omega} \frac{\left(\left(1+\frac{\beta_2 m\omega\Omega_e}{\alpha}\right)\right)^{-1}\left(1+\frac{1}{4}(m\omega)^2\tau_\rho^2 L_0 N_s\right)N_s\pi\tau^2\Xi_m}{\left(\frac{1}{2}(m\omega)^2\tau_\rho^2 L_0 + \frac{2}{N_s} + \alpha L_0\right)\left(1+\frac{N_s}{2}\beta_2 m\omega L_0 \Omega_e + \frac{N_s}{4}(m\omega)^2\tau_\rho^2 L_0\right)}. \quad (22)$$

Table I lists the roll-offs of its single terms for the test bed conditions (Section III A.). For large numbers of spans, it simplifies and enables a separation of the NLDP trends as a function of system parameters

$$\sigma_{NLDP\atop m}^2(L_e,0) - \sigma_{NLDP\atop m}^2(L_e,\tau) =$$
$$\frac{2\gamma^2 e^{-\alpha L_0}\Omega_e^2}{9\beta_2 m\omega\alpha\omega} \frac{\left(1+\frac{1}{4}(m\omega)^2\tau_\rho^2 L_0 N_s\right)N_s\pi\tau^2\Xi_m}{\left(\frac{2}{N_s}+\alpha L_0\right)\left(1+\frac{N_s}{2}\beta_2 m\omega L_0\Omega_e\right)\left(1+\frac{\beta_2 m\omega\Omega_e}{\alpha}\right)} \quad (23)$$
$$\approx \left(\frac{\gamma}{\alpha\beta_2 L_0}\right)^2 \frac{4}{9} \frac{e^{-\alpha L_0}\Omega_e}{(m\omega)^2\omega} \frac{\left(1+\frac{1}{4}(m\omega)^2\tau_\rho^2 L_0 N_s\right)\pi\tau^2\Xi_m}{\left(1+\frac{\beta_2 m\omega\Omega_e}{\alpha}\right)}$$

Summation over '$m$' while assuming flat spectra as shown in Fig.2a yields

$$\sigma_{NLDP}^2(L_e,0) - \sigma_{NLDP}^2(L_e,\tau)$$
$$\approx \frac{8}{9}\left(\frac{\gamma}{\alpha\beta_2 L_0}\right)^2 \frac{e^{-\alpha L_0}\Omega_e}{\omega^2}\pi\tau^2\Xi_m \left[\frac{\beta_2\Omega_e}{\alpha}ln\left(\frac{1+\frac{\beta_2\Omega_e}{\alpha}\Omega_{max}}{\Omega_{max}}\Omega_{min}\right) + \frac{1}{\Omega_{min}} + \frac{1}{4\beta_2\Omega_e}\tau_\rho^2 L_0 N_s ln\left(1+\frac{\beta_2\Omega_e}{\alpha}\Omega_{max}\right)\right]. (24)$$

Next, we compare our model's prediction with the standard deviation of the SOP speed from the 10 Mm transmission plotted in Fig.2c

$$\sqrt{\langle SOP_{NLDP}^2\rangle} = \sqrt{\left\langle \frac{\left(\overline{S(\tau_s+t)} - \overline{S(t)}\right)^2}{\tau_s^2}\right\rangle}, \quad (25)$$

where $\vec{S}$ and $\tau_s$ denotes the probe's Stokes vector and the sampling time, respectively. SOP speed computation requires converting the perturbation from Jones into Stokes space. After some amount of algebra that transforms the differential (Eq. (24)) into Stokes space and after its averaging across the Poincare sphere one finds

$$\langle SOP_{NLDP}^2\rangle \approx$$
$$\frac{20}{27}\frac{\gamma^2}{\beta_2^2\alpha L_0}\left(\frac{\Omega_e}{\alpha L_0}\frac{\pi}{\Omega_{min}} + \frac{\tau_\rho^2}{4}\frac{1}{\beta_2}N_s\pi\ln\left[1+\frac{\beta_2\Omega_e}{\alpha}\Omega_{max}\right]\right)\left(\frac{P_{Rep}}{(\Omega_{max}-\Omega_{min})}\right)^2. (25)$$

Its computed value $\sqrt{<SOP_{NLDP}^2>}\sim 8.4$ Mrad/s diverges from the experimental amount ~3.8 Mrad/s (processed from Fig.2c) and we leave improving the several approximations made during its derivation to future work. Also, our model does not include PDL effects that need to be considered when discussing a $N_s$-independent offset of Eq. (25) also visible in Fig.2c.

TABLE I. Roll-Off frequencies of terms in (22)

| | simplification | 3-dB roll-off frequency | Value [THz] |
|---|---|---|---|
| $1+\frac{1}{4}(m\omega)^2\tau_\rho^2 L_0 N_s$ | | $\frac{1}{\pi\tau_\rho\sqrt{L_0}\sqrt{N_s}}$ | 0.08 |
| $\frac{1}{2}(m\omega)^2\tau_\rho^2 L_0 + \frac{2}{N_s} + \alpha L_0$ | $\frac{2}{N_s} \ll \alpha L_0$ | $\frac{1}{\pi\sqrt{2}}\frac{\sqrt{\alpha}}{\tau_\rho}$ | 1 |
| $1 + \frac{N_s}{2}\beta_2 m\omega L_0 \Omega_e + \frac{N_s}{4}(m\omega)^2\tau_\rho^2 L_0$ | $\beta_2\Omega_e \gg \frac{1}{2}m\omega\tau_\rho^2$ | $\frac{1}{\pi N_s \beta_2 \Omega_e L_0}$ | 0.007 |
| $1 + \frac{\beta_2 m\omega \Omega_e}{\alpha}$ | | $\frac{\alpha}{2\pi\beta_2\Omega_e}$ | 1.1 |

## V. NLDP Impact on the System Channel Capacity

Modern optical communication systems pol-mux two orthogonal channels at same wavelength to maximize the spectral efficiency. Their instantaneous common receive SOP is equally blurred by NLDP in both azimuth and polar angles on the Poincare sphere. Especially, fast SOP motions in azimuthal direction (assuming the individual channels possess polarizations aligned with the x-y coordinates) cause significant coherent crosstalk at high receive OSNR for advanced modulation formats and reduce established limits for the channel capacity. Current theories for capacity limits have not explicitly considered NLDP. They are based on the Manakov equation[20,21,22,23,24,25,26,27,28] or its simplified versions whose derivation for birefringent fiber includes an SOP averaging[29,30] over the Poincare sphere that needs to be revisited for NLDP applications. We note, the Manakov equation does not include a parameter for the fiber PMD, which however is essential for the NLDP magnitude (see e.g. Eq.(24)). It also assumes different magnitudes for the fields that cause cross-phase modulation. Therefore, without incorporating additional assumptions into the fiber model, the applicability of the Manakov equation to NLDP is limited. In the Manakov-PMD equation, which is generally considered to cover all known propagation effects, the NL PMD term vanishes for sufficiently strong varying fiber birefringence as it is the case for fibers in our setups[18]. Thus, the origin of NLDP should not be confused with NL PMD.

NLDP accumulates over wide bandwidth and long reach which complicates a split-step method-based simulation[31] in terms of required computational resources. Typically, such simulation techniques for system design assume simplified models for cross-phase modulation between a carrier and parts of a spectrum that are widely spaced to reduce the computational effort. These techniques however, do not accurately represent NLDP.

## VI. Conclusion

Unpolarized optical noise slightly depolarizes light in commercial long-haul communication systems and lab test-beds. Although comparable small to other polarization effects, this phenomenon leads to a qualitatively different microscopic understanding of nonlinear light propagation in fiber. We have demonstrated by means of SOP speed histograms the dependence of NLDP on the transmission distance and the repeater output power in long-haul links. Predictions from a theory for NLDP based on the coupled nonlinear Schrodinger equations qualitatively agree with experimental observations. Reassessing under consideration of NLDP fiber channel capacity simulations that are utilizing Manakov type equations can be beneficial for scientific purposes and could show small performance offsets.